\documentstyle[aps,twocolumn,eqsecnum,epsfig]{revtex}

\def\Slash#1{{\ooalign{\hfil$#1$\hfil\crcr\hfil$/$\hfil}}}

\begin{document} 
\draft
\preprint{TU-624}
\title{Vacuum structure of Toroidal Carbon Nanotubes}
\author{K. Sasaki}
\address{
Department of Physics, Tohoku University, Sendai 980-8578, Japan
}
\maketitle
\abstract{
Low energy excitations in carbon nanotubes can be 
described by an effective field theory of two 
components spinor.
It is pointed out that the chiral anomaly in 
1+1 dimensions should be observed in a
metallic toroidal carbon nanotube 
on a planar geometry with varying magnetic field. 
We propose an experimental setup for studying this quantum effect.
We also analyze the vacuum structure of the metallic toroidal carbon nanotube 
including the Coulomb interactions and discuss some effects 
of external charges on the vacuum.}
\section{introduction}

In recent years carbon nanotubes(CNTs)~\cite{Iijima}
have attracted much attention from various points of view.
Especially their unique mechanical and electrical properties have
stimulated many people's interest in the analysis of CNTs~\cite{RS,KT,Dek}. 
They have exceptional strength and stability,
and they can exhibit either metallic or semiconducting depending on the
diameter and helicity~\cite{JWM,WV}.
Because of their small size, properties of CNTs should be 
governed by the law of quantum mechanics. 
Therefore it is quite important to understand the 
quantum behavior of the electrons on CNTs. 
The bulk electric properties of (single-walled)CNTs are
relatively simple,  
but the behavior of electrons at the end of a tube(cap) or 
metal-CNT junction is complicated
and its understanding is necessary for building actual 
electrical devices. On the other hands, toroidal CNTs 
(Fullerence `Crop Circles'~\cite{cc}) are clearly simple
because of their no-boundary shape and they can also be either metal or 
semiconducting properties( hereafter we use `torus' or `nanotorus' instead of 
`toroidal carbon nanotube' for simplicity).
Even in the torus case, quite important effect: 
`chiral anomaly'~\cite{Bell},
which is of essentially quantum nature, might occur.

Low energy excitations on CNTs at half filling move along 
the tubule axis because the circumference degree of freedom
(an excitation in the compactified direction) 
is frozen by a wide energy gap. 
Hence this system can be described as a 1+1 dimensional system.
Furthermore in the case of metallic CNTs, the system describing
small fluctuations around the Fermi point is equivalent to the
massless fermion in 1+1 dimensions. If we include gauge 
field, this situation can be modeled by the
quantum field theory of massless fermion which couples to 
the gauge field through minimal coupling.
This model realizes the chiral anomaly phenomenon~\cite{Sch,Man,Iso}.

The chiral anomaly is one of the most interesting phenomena in 
quantum field theory and has had an appreciable influence on the modern
development of high energy physics~\cite{Jack} and of condensed matter physics~\cite{NN}. 
The effect of the chiral anomaly on the electrons in a nanotorus appears
directly as a current flow.
On the other hand, 
it is known in solid state physics that a one-dimensional metallic ring shows 
the `persistent current~\cite{persist}' in an appropriate experimental setting.
The current originating from the chiral anomaly shows the same magnetic field dependence 
to the persistent current. Therefore, the persistent current is closely connected with 
the chiral anomaly in 1+1 dimensions.
The chiral anomaly provides a deeper understanding for the persistent current as 
shown in the present paper.

In this paper, 
we examine the anomaly effect in a metallic nanotorus and
discuss how such an effect can be observed experimentally.
We also clarify the vacuum structure of 
the model regarding the gauge field
as a classical field and discuss some effects of external charges
situated on the metallic torus.

The organization of this paper is as follows.
After reviewing quantum physics of CNTs~\cite{JG}, 
we study the case of nanotorus in section \ref{sec:cnt-torus}.
In section \ref{sec:eft-cnt} we point out that low energy excitations
on a metallic torus at half filling can be modeled by a
quantum field theory of massless fermions with gauge field
and construct the Hamiltonian of this system.
In section \ref{sec:ca-cnt-torus}
we discuss the chiral anomaly and show how such an effect can be observed
experimentally.
We examine the Hamiltonian including the Coulomb interactions and 
analyze an effect of the Coulomb interactions on the chiral anomaly 
in section \ref{sec:vac-cnt-torus}.
We discuss some effects of external charges 
in section \ref{sec:eff-external}.
Conclusion and discussion are given in section \ref{sec:discuss}.

\section{carbon nanotorus}\label{sec:cnt-torus}

\begin{figure}
\begin{center}
\includegraphics[scale=0.3]{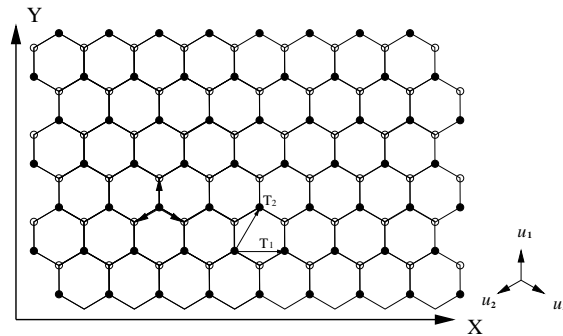}
\end{center}
\caption{Lattice structure of a two-dimensional graphite sheet
($ u_1 = a e_y,\ u_2 = -\frac{\sqrt{3}}{2}a e_x - \frac{1}{2}a e_y,\
u_3 = \frac{\sqrt{3}}{2}a e_x - \frac{1}{2}a e_y$)}
\label{fig:honeycomb}
\end{figure}

A carbon nanotube can be thought of as a layer of graphite 
sheet folded-up into a cylinder. 
A Graphite sheet consists of many hexagons 
whose vertices are occupied
by the carbon atoms and each carbon supplies one conducting electron 
which determines the electric properties of the graphite sheet. 
The lattice structure of a two-dimensional graphite sheet is shown
in Fig.\ref{fig:honeycomb}. 
It is obvious, however, that this picture of a CNT as a graphite sheet
rolled up to form a compact cylinder is somewhat oversimplified.
We need to be careful of the following facts.
First, a conducting electron makes the
$\pi$ orbitals whose wave function extends into the Z-direction:
perpendicular to the graphite sheet. 
Hence, in the case of multi-walled CNTs(MWCNTs), 
the wave functions which belong to 
different layers may interfere so that there is a chance that 
some electrical properties will alter~\cite{Multi} as compared with SWCNTs. 
Second, we should care for the curvature of a
cylinder. This cause the mixing of $\sigma$ and $\pi$ orbital so that 
the band structure might change.
In the present paper, we concentrate single-walled CNTs 
with a large diameter.
Therefore the effects of interlayer interactions and curvature 
can safely be neglected~\cite{Kim}.

There are two symmetry translation vectors on the planar honeycomb lattice,
\begin{eqnarray}
T_1 = \sqrt{3} a e_x , \ \ \ T_2 = \frac{\sqrt{3}}{2} a e_x + \frac{3}{2} a e_y.
\end{eqnarray}
Here $a$ denotes the length of the nearest carbon vertex and is given by 0.142 nm.
$e_x$ and $e_y $ are unit vectors which are orthogonal to each other 
($e_x \cdot e_y = 0$).
If we neglect the spin degrees of freedom,
because of these symmetry translations, the Hilbert space is spanned 
by the following two Bloch basis vectors,
\begin{eqnarray}
&&
|\Psi^k_\bullet \rangle = \sum_{i \in \bullet} e^{ikr_i} a^\dagger_i |0 \rangle,
\\
&&
|\Psi^k_\circ \rangle = \sum_{i \in \circ} e^{ikr_i} a^\dagger_i |0 \rangle ,
\end{eqnarray}
where the black($ \bullet $) and blank($ \circ $) 
indices are indicated in Fig.\ref{fig:honeycomb}.
$r_i $ labels the vector pointing each site $i $, and 
$a_i,a_j^\dagger $ are canonically annihilation-creation operators of 
the electrons of site $i$ and $j$ that satisfy 
\begin{eqnarray}
\{ a_i,a_j^\dagger \} = \delta_{ij} .
\end{eqnarray}
We construct a state vector which is an eigenvector of these
symmetry translations as follows:
\begin{eqnarray}
|\Psi^k \rangle = C^k_\bullet |\Psi^k_\bullet \rangle 
+ C^k_\circ |\Psi^k_\circ \rangle.
\end{eqnarray}
In order to define the unit cell of wave vector $k$, 
we act the symmetry translation operators on the state vector and 
obtain the Brillouin zone
\begin{eqnarray}
-\frac{\pi}{\sqrt{3}} \le a k_x < \frac{\pi}{\sqrt{3}},\
-\pi \le \frac{\sqrt{3}}{2}a k_x + \frac{3}{2}a k_y < \pi,
\end{eqnarray}
where $k_x = k \cdot e_x $.

The mapping of the graphite sheet onto a cylindrical surface is
specified by a wrapping vector: 
\begin{eqnarray}
C = N T_1 + M T_2 ,
\label{eq:chiral vector}
\end{eqnarray}
which defines the relative location of the two sites.
The pair of indices $(N,M)$ describes how the sheet is wrapped to 
form the cylinder, 
and determines its electrical properties~\cite{JWM,WV}.
It can be divided CNTs into three categories depending on 
the pair of integers $(N,M)$.
A tube is called `zigzag' type if $M=0$ and `armchair' in the
case $M=-2N$. All other tubes are of the `chiral' type.

To study the electronic properties of a nanotorus quantum mechanically, 
first we compactify the graphite sheet into a cylinder by imposing a
periodic boundary condition to the state vector.
In general, we may consider the following boundary condition,
\begin{eqnarray}
\hat{G}(NT_1 + MT_2) |\Psi^k \rangle = |\Psi^k \rangle.
\end{eqnarray}
$\hat{G} $ denotes a symmetry translation operator.
From which we obtain
\begin{eqnarray}
\frac{\sqrt{3}}{2}\left( 2N + M \right)a k_x  
+ \frac{3}{2} M a k_y = 2\pi n ,
\label{eq:spectrum lines}
\end{eqnarray}
where n is an integer.
Next, in oder to make a torus, 
we compactify the tube into a torus by imposing 
a boundary condition to the tubule axis direction.
For example, we may consider a `zigzag torus' which has the following 
boundary conditions,
\begin{eqnarray}
&&
\hat{G}(NT_1) |\Psi^k \rangle = |\Psi^k \rangle ,
\nonumber
\\
&&
\hat{G}(M(2T_2-T_1)) |\Psi^k \rangle = |\Psi^k \rangle.
\label{eq:pbc}
\end{eqnarray}
The former condition makes a sheet into a zigzag tube, 
and the latter forces the tube into a zigzag torus.

\begin{figure}
\begin{center}
\includegraphics[scale=0.3]{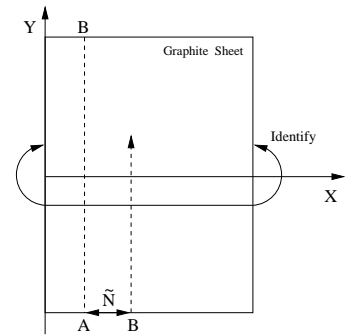}
\end{center}
\caption{Twisted zigzag torus: We impose the periodic boundary condition
in the X-direction and attack same character(`B') in the Y-direction to
obtain a twisted nanotorus.
There are $\tilde{N}$ hexagons between the point A and B.}
\label{fig:twist}
\end{figure}

It is clear that there are many possibilities for the shape of torus
and each shape has its own boundary condition. So,
some of them have different properties from the one (\ref{eq:pbc}). 
Especially we can image a torus in which some
twist exists along the tubule axis direction~\cite{CBJ}.
This system have the following boundary conditions,
\begin{eqnarray}
&&
\hat{G}(NT_1) |\Psi^k \rangle = |\Psi^k \rangle ,
\nonumber
\\
&&
\hat{G}(M(2T_2-T_1))|\Psi^k \rangle = 
\hat{G}(\tilde{N}T_1) |\Psi^k \rangle ,
\label{eq:tbc}
\end{eqnarray}
where $\tilde{N} $ is decided by the twist at the junction of tube end, 
see Fig.\ref{fig:twist}.
These boundary conditions yield the discrete wave vectors
\begin{eqnarray}
a k_x = \frac{2\pi}{\sqrt{3}} \frac{n}{N}, \ \
a k_y = \frac{2\pi}{3M} \left( m+ \frac{\tilde{N}}{N} \right).
\label{eq:discretekx}
\end{eqnarray}

So far we have constructed the Hilbert space of the conducting
electrons. The Hilbert space is spanned by the Bloch basis vectors
with the discrete wave vectors (\ref{eq:discretekx}).
Now, we consider the Hamiltonian which governs 
the time evolution of a state vector.
Each carbon atom has an electron which makes $\pi $-orbital. 
The electron can transfer from any site to the nearest 
three sites through the quantum mechanical 
tunneling or thermal hopping in finite temperature. 
Therefore there is some probability amplitude for this process.
In this case, the tight-binding Hamiltonian is most suitable:
\begin{eqnarray}
{\cal H} = E_0 \sum_i a^\dagger_i a_i 
+ \gamma \sum_{\langle i,j \rangle} a_i^\dagger a_j,
\end{eqnarray}
where the sum $\langle i,j \rangle $ is over pairs of 
nearest-neighbors carbon atoms $i,j$ on the lattice. 
$\gamma $ is the transition amplitude from 
one site to the nearest sites and $E_0 $ is the one from a site 
to the same site. This parameter $E_0 $ only fixes the origin of the 
energy and therefore irrelevant.
It is an easy task to find the energy eigenstates and eigenvalues 
of this Hamiltonian.
In the matrix representation, the energy eigenvalue equation reads
\begin{eqnarray}
\pmatrix{ E_0 & \gamma \sum_i e^{iku_i} \cr
\gamma \sum_i e^{-iku_i} & E_0 \cr}
\pmatrix{ C_\bullet^k \cr C_\circ^k}
= E_k \pmatrix{ C_\bullet^k \cr C_\circ^k},
\end{eqnarray}
where
\begin{eqnarray}
|\Psi^k_\bullet \rangle 
= \pmatrix{ 1 \cr 0 \cr},\ \ |\Psi^k_\circ \rangle 
= \pmatrix{ 0 \cr 1 \cr},
\end{eqnarray}
and the vector $u_i $ is a triad of vectors pointing
respectively in the direction of the nearest neighbors of a black
point, see Fig.\ref{fig:honeycomb}. 
The energy eigenvalues and eigenvectors are as follows,
\begin{eqnarray}
&& E_k = E_0 \pm \Delta(k), \\
&& \pmatrix{ C_\bullet^k \cr C_\circ^k}
= \frac{1}{\sqrt{2}\Delta(k)}
\pmatrix{ \gamma \sum_i e^{iku_i} \cr \pm \Delta(k)},
\end{eqnarray}
where
\begin{eqnarray}
\Delta(k)= \gamma \sqrt{1+ 4 \cos{\frac{\sqrt{3}}{2}a k_x} 
\cos{\frac{3}{2}a k_y} + 4 \cos^2{\frac{\sqrt{3}}{2}a k_x}}.\nonumber
\end{eqnarray}
The structure of this energy band has striking properties
when considered at half filling.
This is the situation which is physically interesting. Since 
each level of the band may accommodate two states due to the 
spin degeneracy, the Fermi level turns out to be at midpoint 
of the band($E_k = E_0 $).
Fermi points in the Brillouin zone are located at
\begin{eqnarray}
a\tilde{k}_{1,2} =
(a k_x , a k_y) = ( \pm \frac{2\pi}{3\sqrt{3}} , \mp \frac{2\pi}{3}) .
\label{2fermi-points}
\end{eqnarray}
Hence, if $N $ in eq.(\ref{eq:discretekx}) 
is a multiple of 3, then the zigzag torus shows metallic properties.

In order to understand the electric properties
we should take into account a small perturbation around the Fermi point. 
So we take $k = \tilde{k}_1 + \delta k$
as a small fluctuation~\cite{note}.
Perturbation around the point $\tilde{k}_2 $ is 
same as around the point $k = \tilde{k}_1 $. 
So we may only consider one of the pairs.
In this case the Hamiltonian which describes the system 
is given by
\begin{eqnarray}
-\frac{3}{2}\gamma \sigma \cdot \delta k
= -\frac{3}{2}\gamma a 
\pmatrix{ 0 & \delta k_x -i \delta k_y \cr
\delta k_x +i \delta k_y & 0 \cr}. 
\label{eq:pert}
\end{eqnarray}
This implies that the low energy excitations of metallic CNTs 
at half filling are described by an effective theory of two dimensional 
spinor obeying the Weyl equation. 
In coordinate space, we find
\begin{eqnarray}
{\cal H}_{pert} = -\frac{3\gamma a}{2\hbar} \sigma \cdot p ,
\end{eqnarray}
where $p$ is the momentum operator, $p = -i \hbar \nabla $ and 
$\sigma_i $ are the Pauli matrices.
It is also convenient to use a parameter 
$\beta = -\frac{3\gamma a}{2\hbar}$ and $T = \beta t $.
In this case the Schr\"odinger equation becomes~\cite{weyl}
\begin{eqnarray}
i\hbar \frac{\partial}{\partial T}\psi 
=\left( \sigma \cdot p \right) \psi.
\end{eqnarray}
In the following section, 
we consider metallic and semi-conducting zigzag 
torus that have small N and large M values($\frac{M}{N} \ge 10^2 $).
In this case, 
transitions between different $k_x$ are very small because of 
their costed energy($\sim \frac{|\gamma|}{N}$) as compared to that 
of $k_y$:($\sim \frac{|\gamma|}{M}$).
Therefore, the only surviving degree is a motion in the tubule axis 
direction, i.e. this system is a 1+1 dimensional system effectively.

\section{effective field theory of carbon nanotube}\label{sec:eft-cnt}

In this section we would like to focus on the zigzag torus 
which has the boundary conditions (\ref{eq:pbc}) and 
construct an effective field theory describing 
the low energy excitations in the torus.
More general boundary conditions(\ref{eq:tbc}) are discussed 
in section \ref{sec:ca-cnt-torus}.
The zigzag torus can exhibit either metallic or semi-conducting 
depending on the value of $N$. 
If $N$ is a multiple of 3 then the torus shows metallic properties.
In order to analyze the semi-conducting case equally,
we set $N = \pm 3n + b $, where $n$ is a positive integer and 
$b \in \{ 0,\pm 1,\pm 2 \}$.
To examine the low energy excitations,
we should consider the following wave vectors and energy: 
\begin{eqnarray}
&& ak_x = \pm \frac{2\pi}{3\sqrt{3}}\frac{1}{1\pm \frac{b}{3n}}, \ \ \
ak_y = \mp \frac{2\pi}{3} + a \delta k_y, \\
&& E_k = E_k \left( 
\pm \frac{2\pi}{3\sqrt{3}}\frac{1}{1\pm \frac{b}{3n}},
\mp \frac{2\pi}{3} + a \delta k_y \right).
\end{eqnarray}
Considering the small perturbation ($a \delta k_y < \frac{2\pi}{9} $) 
and a large diameter ($ n \ge 3$), 
the excitation energy can be approximated by
\begin{eqnarray}
E_k \sim \pm v_F
\sqrt{M^2 + p^2},
\end{eqnarray}
where $v_F(\equiv |\beta|) $ is the Fermi velocity, $p = \hbar \delta k_y$ and 
\begin{eqnarray}
M = \frac{2\pi \hbar}{\sqrt{3}3na}\frac{|b|}{3}  .
\end{eqnarray}
Consequently, we have obtained a linear dispersion relation for 
metallic case($b=0$). This is to be contrasted with the dispersion
relation for semi-conducting case($b\ne 0$).

Here we comment on the excitations between different $k_x$.
In order to neglect these excitations, 
we have to restrict the energy to the following region:
\begin{eqnarray}
E_k < E_{\delta k_x} = \frac{\pi}{\sqrt{3}N} \gamma .
\end{eqnarray}
At room temperature, thermal excitation between different $|\delta k_x|$
cannot occur by the Boltzmann suppression factor,
$e^{-\frac{E_{\delta k_x}}{kT}} (\sim e^{-\frac{200}{N}})$,
so that the effects of different $|\delta k_x|$ can be ignored.
Accordingly, 
the Hamiltonian which describes the low energy 
excitations near the Fermi point is given by
\begin{eqnarray}
-\frac{3}{2}\frac{\gamma a}{\hbar} 
\pmatrix{ 0 & \pm M -i p \cr
\pm M +i p & 0 \cr}.
\end{eqnarray}
The sign $(\pm)$ in front of $M$ comes from the 
$\tilde{k}_1$ and $\tilde{k}_2$ points.
The sign can be removed by an appropriate unitary transformation of 
the state vector. Therefore we may choose the minus sign without loss
of generality.
Using the unitary transformation,
$U =e^{-i\frac{\pi}{4} \sigma_x}$,
the Schr\"odinger equation becomes
\begin{eqnarray}
i\hbar \frac{\partial }{\partial T} \psi = 
\pmatrix{ p &  -M \cr -M & -p } \psi.
\label{eq:Schr}
\end{eqnarray}

One can obtain the quantum field theory by promoting
the wave function($\psi$) to the field operator($\Psi$) obeying the
anti-commutation relation.
Because the Schr\"odinger equation (\ref{eq:Schr}) is the Dirac 
equation in two dimensions, 
it is appropriate to adopt the following Lagrangian density. 
Addition of the electro-magnetic interaction according to the 
minimal coupling gives
\begin{eqnarray}
{\cal L} = -\frac{1}{4}F^{\mu\nu}F_{\mu\nu} 
- \bar{\Psi} \left( \Slash{D} + M \right) \Psi ,
\label{eq:lag}
\end{eqnarray}
where $D$ is the covariant derivative and $F_{\mu\nu}$ is 
the field strength,
\begin{eqnarray}
&&
\Slash{D} = \sum_{\mu =0,1}
\left(i\hbar \partial_\mu - \frac{e}{c} A_\mu \right) 
\gamma^\mu, \nonumber \\
&&
F_{\mu\nu} = \partial_\mu A_\nu - \partial_\nu A_\mu, \ \ \bar{\Psi} = \Psi^\dagger \gamma^0.
\end{eqnarray}
The gauge fields($A_\mu$) live in four dimensional space-time so that 
summention indices run from 0 to 3 in the gauge kinetic term.
We adopt the following notation:
\begin{eqnarray}
&&
g^{\mu\nu} = diag\{ 1,-1,-1,-1\},\ \ 
\gamma^0 = \pmatrix{ 0 & 1 \cr 1 & 0}, \nonumber
\\
&&
\gamma^1 = \pmatrix{ 0 & 1 \cr -1 & 0},\ \
\gamma^5 =-\gamma^0 \gamma^1 =  \pmatrix{ 1 & 0 \cr 0 & -1}.
\end{eqnarray}

We quantize the fermion field in each configuration 
of the gauge field choosing Weyl gauge condition ($A_0 =0$). 
Fermionic part of the Hamiltonian density is given by
\begin{eqnarray}
{\cal H} = {\cal H}_F + {\cal H}_C .
\end{eqnarray}
The total Hamiltonian density consists of the kinetic term and the 
Coulomb term. The kinetic term is given by
\begin{eqnarray}
{\cal H}_F &&= \Psi^\dagger h_F \Psi \nonumber \\
&&
= \Psi^\dagger v_F 
\pmatrix{ i\hbar \partial_1 -\frac{e}{c} A_1 & -M \cr
                  -M & -( i\hbar \partial_1 -\frac{e}{c} A_1) } \Psi.
\end{eqnarray}
The Coulomb interaction term is
\begin{eqnarray}
{\cal H}_C = \frac{e^2}{8\pi} \int \frac{J^0(x)J^0(y)}{|x-y|} dy .
\end{eqnarray}
where $e J^0(x)$ stands for the charge density.
The electric current is given by $e J^\mu = e \bar{\Psi} \gamma^\mu \Psi$
where $e$ is the electron charge. 

It should be note that besides the Coulomb interaction, 
backscattering and umklapp process may come in the dynamics 
of electrons as is shown in reference~\cite{EG}.
We neglect these interactions providing that their coupling are weak enough.

\section{chiral anomaly in a metallic nanotorus}\label{sec:ca-cnt-torus}

We have obtained the Hamiltonian density
which describes the low energy excitations in the zigzag torus. 
The Hamiltonian consists of two parts, one is the kinetic term and
the other is the Coulomb interaction.
In this section we discuss the quantum mechanical vacuum structure
of the kinetic Hamiltonian: $H_F (=\oint {\cal H}_F)$. 
Effects of the Coulomb interaction will be considered in latter sections.
Hereafter we focus on the metallic case and set $v_F =1 $ in this section 
for simplicity.
In the metallic case: $b=0$, the energy eigenvectors are given by
\begin{eqnarray}
&&
h_F \psi_n \pmatrix{ 1 \cr 0 } = \epsilon_n \psi_n \pmatrix{ 1 \cr 0 },\
h_F \psi_n \pmatrix{ 0 \cr 1 } = -\epsilon_n \psi_n \pmatrix{ 0 \cr 1 },
\nonumber
\\
&&
\ \ \ \psi_n(x) = \frac{1}{\sqrt{L}} e^{-i\frac{e}{\hbar c} 
\int_0^x A_1(y) dy -i\frac{\epsilon_n}{\hbar} x},
\end{eqnarray}
where $\epsilon_n$ is the energy eigenvalues and 
$L$ is the circumferential length of the zigzag torus: $L=3a|M|$.
We expand the fermion field using the left and right moving waves as
\begin{eqnarray}
&&\Psi(x,T)= \Psi_L(x,T) + \Psi_R(x,T) \nonumber \\
&&= \sum_{n\in Z} 
\left[ a_n \psi_n(x) e^{-i\frac{\epsilon_n}{\hbar}T}
\pmatrix{ 1 \cr 0} + b_n \psi_n(x) e^{+i\frac{\epsilon_n}{\hbar}T}
\pmatrix{ 0 \cr 1} \right] ,
\end{eqnarray}
where $a_n,b_n$ are independent fermionic annihilation operators 
satisfying the anti-commutators:
\begin{eqnarray}
\{ a_n , a^\dagger_m \} = \{ b_n , b^\dagger_m \} = \delta_{nm} .
\end{eqnarray}
All the other anticommutator vanish.

\begin{figure}
\begin{center}
\includegraphics[scale=0.35]{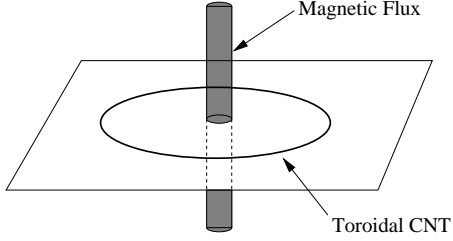}
\end{center}
\caption{A toroidal carbon nanotube on a planar geometry with magnetic field}
\label{fig:setup}
\end{figure}

In order to get the energy spectrum $\epsilon_n$,
we have to impose a boundary condition to the eigenfunctions.
We take (\ref{eq:tbc}) as a general boundary condition.
The boundary condition of the zigzag torus in the 
Y-direction becomes the following:
\begin{eqnarray}
\hat{G}(M(2T_2-T_1))|\Psi^k \rangle = 
e^{\pm i\frac{2\pi}{3}\tilde{N}} |\Psi^k \rangle ,
\end{eqnarray}
where plus(minus) in exponent has its origin in 
the Fermi point, $\tilde{k}_1(\tilde{k}_2)$.
Hence we should impose the following boundary conditions on the 
fermion energy eigenfunctions,
\begin{eqnarray}
\psi_n( x + L) = e^{\pm i\frac{2\pi}{3}\tilde{N}} \psi_n(x) ,
\end{eqnarray}
then the energy eigenvalues are given by
\begin{eqnarray}
\epsilon_n = \frac{\hbar}{L}
\left[ 2\pi (n \pm \frac{\tilde{N}}{3}) -
\frac{e}{\hbar c} \oint_0^L A_1 dy \right].
\end{eqnarray}
Here, the gauge field $A_1$ is experimentally controllable 
by the following experimental setup, see Fig.\ref{fig:setup}.
On a planar geometry we set a nanotorus
and put some magnetic field inside the torus perpendicular
to the plane.
In this case the gauge field that expresses this 
magnetic field is given by, in the vector notation,
$A = \frac{N\phi_D}{2\pi} \nabla \theta$
where $\theta$ is an angle of two points on the torus.
Therefore we get a component,
\begin{eqnarray}
A_1 = \frac{N \phi_D}{L},
\end{eqnarray}
where $\phi_D = \frac{2\pi \hbar c}{e} $. 
This vector potential expresses the N flux inside the torus and 
by tuning the magnetic field, N can be taken as a real number.

The Lagrangian density (\ref{eq:lag}) with massless fermion 
has two classical conserved currents
\begin{eqnarray}
&&
J^\mu(x) = \bar{\Psi}(x) \gamma^\mu \Psi(x),
\\
&&
J^\mu_5(x) = \bar{\Psi}(x) \gamma^\mu \gamma^5 \Psi(x) 
= \epsilon^{\mu\nu}J_\nu(x),
\label{eq:current}
\end{eqnarray}
where $\epsilon^{\mu\nu}$ is an antisymmetric tensor: $\epsilon^{01} = 1$.
Therefore, the following two charges conserve in the time evolution of the
system at the classical level,
\begin{eqnarray}
&&
Q = \oint J^0(x)dx , 
\\
&&
Q_5 = \oint J^0_5(x)dx .
\end{eqnarray}
$J^\mu$ (electric current) conservation($\partial_\mu J^\mu=0$) is due to the gauge symmetry
and $J^\mu_5 $ (chiral current) conservation($\partial_\mu J_5^\mu=0$) is due to the global
chiral symmetry($\Psi \to e^{i\gamma_5 \alpha} \Psi$).

Different from classical mechanics,
in the world of quantum mechanics, the chiral symmetry is broken~\cite{Bell}.
In order to find what is happening, we need to analyze the vacuum
structure: $|vac;N_L,N_R \rangle = |vac;N_L \rangle \otimes |vac;N_R \rangle$,
where
\begin{eqnarray}
|vac;N_L \rangle = \prod_{n=-\infty}^{N_L -1} a_n^\dagger |0\rangle,\ \ 
|vac;N_R \rangle = \prod^{n=\infty}_{N_R} b_n^\dagger |0\rangle.
\end{eqnarray}
We define $|vac;N_L \rangle(|vac;N_R \rangle ) $ such that 
the levels with energy lower than $\epsilon_{N_L}(-\epsilon_{N_R -1})$ 
are filled and the others are empty.
On this vacuum the charge expectation values and the energy become
\begin{eqnarray}
&&
\langle Q \rangle = N_L - N_R,
\\
&&
\langle Q_5 \rangle = N_L + N_R -2N -1 \pm \frac{2}{3}\tilde{N} ,
\label{eq:chiralcharge}
\\
&&
\langle H_F \rangle = \frac{2\pi \hbar}{L}
\left( \frac{\langle Q \rangle^2 + \langle Q_5 \rangle^2}{4}
-\frac{1}{12} \right).
\label{eq:fermionE}
\end{eqnarray}

To obtain the above results,
we have regularized the divergent eigenvalues on the vacuum by 
$\zeta$-function regularization,
for example, the gauge charge is regularized as follows:
\begin{eqnarray}
Q = \lim_{s\to 0} \left( \sum_{n \in Z} a_n^\dagger a_n 
\frac{1}{|\lambda \epsilon_n|^s}
+ \sum_{n \in Z} b_n^\dagger b_n 
\frac{1}{|-\lambda \epsilon_n|^s} \right),
\end{eqnarray}
where $\lambda$ is an arbitrary constant with dimension 
of length which is necessary to make $\lambda \epsilon_n $ 
dimensionless. This regularization
respects gauge invariance because the energy of each level is a gauge 
invariant quantity.

It can be shown that the gauge charge $\langle Q \rangle$ must vanish 
in order that the state is a physical state. We now have $ N_L = N_R $.
From the above equation (\ref{eq:chiralcharge}), 
it can be seen that if $N_L$ and $N_R$ are conserved, then,
by varying $N$(the magnetic field), the chiral charge also changes.
Therefore it is not a conserved quantity. 
We thus see that the vacuum is responsible for non-conservation of
chirality even though the dynamics is chirally invariant.
From eq.(\ref{eq:current}) we see that 
the chiral current $J^0_5 $ is equivalent to the electric current 
in the tubule axis direction,
\begin{eqnarray}
Q_5 = \oint J^0_5(x)dx = - \oint J^1(x) dx = - J^1.
\end{eqnarray}
Hence, in order to observe the anomaly, we should observe the electrical 
current in the torus.

Due to the existence of the two Fermi points, the total current in the 
torus is given by the sum of two currents $\langle Q_5 \rangle_{\tilde{k}_1}$
and $\langle Q_5 \rangle_{\tilde{k}_2}$.
We should care for the sign($\pm$) 
in the chiral charge (\ref{eq:chiralcharge}).
We define 
\begin{eqnarray}
&&
\langle Q_5 \rangle_{\tilde{k}_1} = N_L + N_R -2N -1 + \frac{2}{3}\tilde{N} ,
\\
&&
\langle Q_5 \rangle_{\tilde{k}_2} = N_L + N_R -2N -1 - \frac{2}{3}\tilde{N} ,
\end{eqnarray}
and treat them separately.

\begin{figure}
\begin{center}
\includegraphics[scale=0.4]{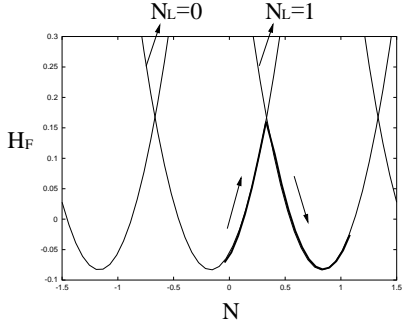}
\end{center}
\caption{Adiabatic change of the fermionic energy of $\tilde{k}_1$ point. 
The energy value is labeled in the unit of $\frac{2\pi \hbar}{L}$}
\label{fig:adiabatic1}
\end{figure}

\begin{figure}
\begin{center}
\includegraphics[scale=0.4]{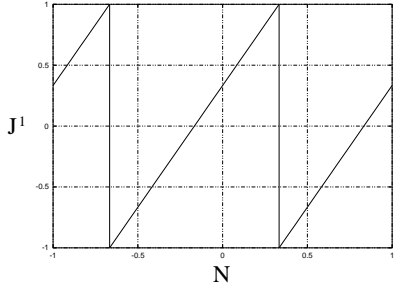}
\end{center}
\caption{Induced current in the twisted zigzag torus: $J^1(\tilde{k}_1) = - \langle Q_5 \rangle_{\tilde{k}_1}$}
\label{fig:current1}
\end{figure}

It is clear from the above equations that there are two origins of the 
usual current flow in the torus. One is the $N_L + N_R$ term 
which can be induced in thermal bath or by a sudden change of the magnetic
field. On the other hand, magnetic field can change the 
quantum vacuum structure and lead to the anomaly.
In order to avoid the unexpected 
changes of $N_L(=N_R) $, the magnetic field must 
be changed adiabatically at low temperature($< \frac{2\pi \hbar}{L}$).
However, in the adiabatic process, when the strength of the
magnetic field reaches the specific points, then $N_L(=N_R)$
also have to change. For simplicity, we set $\tilde{N}=1$ and 
focus on the $\langle Q_5 \rangle_{\tilde{k}_1}$.
When increasing $N$ starting 
from the point $N=0,N_L=0$, 
the energy is going up as in eq(\ref{eq:fermionE}).
At $N = \frac{1}{3}$, the spectrum meets an another line of spectrum 
$N_L=1$, see Fig.\ref{fig:adiabatic1}.
Therefore the circular current for $\tilde{k}_1$ in the ring
\begin{eqnarray}
J^1(\tilde{k}_1) = 2 \left( N-N_L \right) +\frac{1}{3}
\label{eq:J^1}
\end{eqnarray}
follows the line shown in Fig.\ref{fig:current1}.

\begin{figure}
\begin{center}
\includegraphics[scale=0.4]{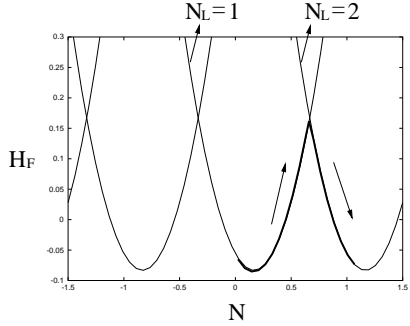}
\end{center}
\caption{Adiabatic change of the fermionic energy of $\tilde{k}_2$ point. 
The energy value is labeled in the unit of $\frac{2\pi \hbar}{L}$}
\label{fig:adiabatic2}
\end{figure}

\begin{figure}
\begin{center}
\includegraphics[scale=0.4]{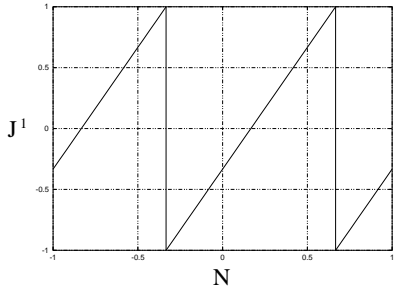}
\end{center}
\caption{Induced current in the twisted zigzag torus: $J^1(\tilde{k}_2) = - \langle Q_5 \rangle_{\tilde{k}_2}$}
\label{fig:current2}
\end{figure}

The same analysis can be applied to the case of 
$J^1(\tilde{k}_2) = - \langle Q_5 \rangle_{\tilde{k}_2}$,
the adiabatic change of energy and induced current 
are shown in Fig.\ref{fig:adiabatic2} and Fig.\ref{fig:current2}.

\begin{figure}
\begin{center}
\includegraphics[scale=0.4]{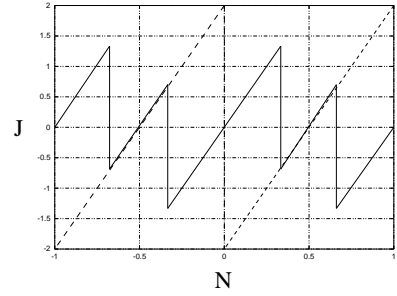}
\end{center}
\caption{Magnetic field dependence of the total current on the torus: $J=J^1(\tilde{k}_1)+J^1(\tilde{k}_2)$}
\label{fig:tcurrent}
\end{figure}

Total current on the torus is given by a sum of two current,
\begin{eqnarray}
J=J^1(\tilde{k}_1)+J^1(\tilde{k}_2).
\end{eqnarray}
Magnetic field dependence of the total current is
shown in Fig.\ref{fig:tcurrent}. 
Dotted line in Fig.\ref{fig:tcurrent} indicates 
the current on the untwisted($\tilde{N} = 0$) torus.
At each Fermi point, there are two spin degrees of freedom.
Therefore the actual current is twice the $J$.
This current for untwisted torus shows the same magnetic field dependence to 
the persistent current in ref.~\cite{persist}.

\section{vacuum structure of a carbon nanotorus}\label{sec:vac-cnt-torus}

In this section we consider the vacuum structure of the total Hamiltonian:
\begin{eqnarray}
&& 
H = \oint {\cal H} = H_F + H_C, \nonumber \\
&&
H_C = \frac{e^2}{8\pi} \oint \frac{J^0(x)J^0(y)}{|x-y|} dx dy .
\label{eq:Coulomb}
\end{eqnarray}
In the previous section we have solved the kinetic part of the total 
Hamiltonian. We clarified its vacuum state and obtained 
the regularized eigenvalues of the physical quantities. 
We saw that the vacuum has the chiral charge
so that it read the chiral anomaly.
What we are interested in at this point is whether the previous results 
change or not by the inclusion of the Coulomb interaction in our 
analysis~\cite{Iso}.
Furthermore we hope to make clear the effects of the 
`external' charge density on the chiral anomaly. 
Their understandings are the first step 
for studying more physically interesting situations 
such as impurity effects and 
junctions of a CNT and a metal or a superconductor.

The Coulomb interaction consists of a product of the charge density.
Therefore it is very convenient to rewrite the kinetic term $H_F$ using
the current operators. For this purpose,
it is useful to introduce the left and right currents as follows:
\begin{eqnarray}
J^0(x) = J_L(x) + J_R(x),
\end{eqnarray}
where 
$J_L(x) = \Psi^\dagger_L(x) \Psi_L(x)$ and 
$J_R(x) = \Psi^\dagger_R(x) \Psi_R(x)$.
We expand these currents by the Fourier modes,
\begin{eqnarray}
J_L(x) = \sum_{n \in Z}  \frac{j_L^n}{L} e^{-i \frac{2\pi nx}{L}},
\
J_R(x) = \sum_{n \in Z}  \frac{j_R^n}{L} e^{+i \frac{2\pi nx}{L}},
\end{eqnarray}
where the Fourier components are the bosonic operators,
\begin{eqnarray}
j_L^n = \sum_{m \in Z} a_m^\dagger a_{m+n},
\ \ \
j_R^n = \sum_{m \in Z} b^\dagger_{m+n} b_m,
\end{eqnarray}
which satisfy the following commutation relations 
on the fermion Fock space,
\begin{eqnarray}
&&
\left[ j_L^n , (j_L^m)^\dagger \right] = n \delta_{n,m},
\\
&&
\left[ j_R^n , (j_R^m)^\dagger \right] = n \delta_{n,m}.
\end{eqnarray}
It is well-known that the following Hamiltonian has the 
same matrix element as the original fermion Hamiltonian:
\begin{eqnarray}
H_F =&&  \frac{2\pi \hbar v_F}{L} 
\bigg\{ \left( \frac{Q^2+Q_5^2}{4} -\frac{1}{12} \right)
\nonumber \\
&& + \sum_{n > 0} \left( (j_L^n)^\dagger j_L^n + (j_R^n)^\dagger j_R^n
\right) \bigg\} ,
\end{eqnarray}
where $v_F$ is the Fermi velocity($v_F \equiv |\beta|$).
This Hamiltonian has a new term 
which is not shown in eq.(\ref{eq:fermionE}).
This term has vanishing value on the previous vacuum state because 
\begin{eqnarray}
j_L^n |vac; N_L \rangle = 0, \ \ \ j_R^n |vac; N_R \rangle = 0,
\end{eqnarray}
for positive $n$. 

The Coulomb interaction can be rewritten using the bosonic current operators,
\begin{eqnarray}
H_C = \frac{e^2}{4\pi L} \sum_{n \ge 0} V(n)
\left( (j_L^n)^\dagger + j_R^n \right)
\left( j_L^n + (j_R^n)^\dagger \right) .
\end{eqnarray}
Here, we introduce the Fourier component of the Coulomb potential,
\begin{eqnarray}
V(n) = 2\pi \int^1_0
dz \frac{\cos (2\pi nz)}{\sqrt{ \sin^2 (\pi z) + \left(\frac{R}{L}\right)^2}} ,
\label{eq:Fourier-coulomb}
\end{eqnarray}
where $R$ is the circumference of the tubule.

Some comments are in order.
When we write the Coulomb interaction in eq.(\ref{eq:Coulomb}), 
it has an ultraviolet divergence in the limit of $x \to y$. 
Therefore we need to introduce some cutoff length. 
It is appropriate that we set it the length of a
diameter of a tubule because we make the approximation that 
mixing of the different momentums in the compactified direction 
cannot happen. This explain the $\left(\frac{R}{L}\right)^2$ term
in the denominator.
Besides, the form of a torus is not a line but 
a ring on a plane so that we should use the 
direct length between $x$ and $y$ in the Coulomb interaction. 
This is the origin of the $\sin^2(\pi z)$ term in the denominator.

We combine the kinetic term and the Coulomb term as follows:
\begin{eqnarray}
H =  H_0 -  \frac{2\pi \hbar v_F}{12L} + \sum_{n > 0} H_n ,
\end{eqnarray}
where
\begin{eqnarray}
H_0 =  \frac{2\pi \hbar v_F}{L} \frac{Q_5^2 + Q^2}{4} + 
\frac{2\pi \hbar c}{L} \frac{\alpha}{4\pi} V(0)
\end{eqnarray}
and 
\begin{eqnarray}
H_n = && \frac{2\pi \hbar v_F }{L} 
\left\{ (j_L^n)^\dagger j_L^n + j_R^n (j_R^n)^\dagger -n \right\} \nonumber
\\
&&
+ \frac{e^2 V(n)}{4\pi L}
\left( (j_L^n)^\dagger + j_R^n \right)
\left( j_L^n + (j_R^n)^\dagger \right) .
\end{eqnarray}
Here we introduce the fine structure constant 
$\alpha(= \frac{e^2}{4\pi \hbar c})$.

\begin{figure}
\begin{center}
\includegraphics[scale=0.5]{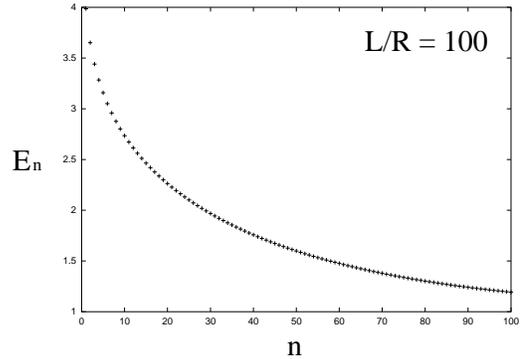}
\end{center}
\caption{Energy spectrums are corrected by the Coulomb interaction. 
The energy value is labeled in the unit of $\frac{2\pi \hbar v_F}{L}$.
Here we set $\frac{L}{R} = 10^2$ in 
eq.(\ref{eq:Fourier-coulomb}).}
\label{fig:ene1}
\end{figure}

We diagonarize the Hamiltonian $H_n (n \ne 0)$
using the Bogoliubov transformation:
\begin{eqnarray}
\pmatrix{ \tilde{j}_L^n \cr (\tilde{j}_R^n)^\dagger }
= \pmatrix{ \cosh t_n  & \sinh t_n \cr \sinh t_n & \cosh t_n }
\pmatrix{ j_L^n \cr (j_R^n)^\dagger },
\end{eqnarray}
where
\begin{eqnarray}
&&
\sinh 2t_n = \frac{1}{E_n} \frac{e^2 V(n)}{4\pi L} ,
\\
&&
\cosh 2t_n = \frac{1}{E_n} 
\left(  \frac{2\pi \hbar v_F}{L} + \frac{e^2 V(n)}{4\pi L} \right) ,
\\
&&
E_n =  \frac{2\pi \hbar v_F}{L} \sqrt{1 + \frac{\alpha}{\pi}\frac{c}{v_F}V(n) }.
\end{eqnarray}
After some calculations we derive
\begin{eqnarray}
H_n = E_n \left[ (\tilde{j}_L^n)^\dagger \tilde{j}_L^n 
+ (\tilde{j}_R^n)^\dagger \tilde{j}_R^n + n \right] - \frac{2\pi \hbar v_F}{L}n .
\end{eqnarray}
The energy $E_n$ differs from $ \frac{2\pi \hbar v_F}{L}$. 
The difference is due to the Coulomb interaction, see Fig.\ref{fig:ene1}.

The generators of the Bogoliubov transformation are given by
\begin{eqnarray}
U_n = \exp \left(-\frac{t_n}{\sqrt{2}n} \{ (\tilde{j}_L^n)^\dagger (\tilde{j}_R^n)^\dagger
- \tilde{j}_L^n  \tilde{j}_R^n \} \right).
\label{eq:BogoTr}
\end{eqnarray}
Hence, we obtain the vacuum state as follows:
\begin{eqnarray}
| \widetilde{vac} \rangle = \left( \prod_{n > 0} U_n \right)
| vac ; N_L,N_R \rangle ,
\end{eqnarray}
where $N_L=N_R$. 
The previous vacuum state changes into new vacuum 
by the Coulomb interaction.
We should estimate the gauge charge and chiral charge on this new vacuum.
Because the operators($U_n$) commute with the charge 
operators $Q$ and $Q_5$, the eigenvalues of these charges do not change.
\begin{eqnarray}
&&
Q |\widetilde{vac} \rangle = 0,
\\
&&
Q_5 |\widetilde{vac} \rangle = 
\left( 2(N_L -N) -1 \pm \frac{\tilde{N}}{3}\right)
|\widetilde{vac} \rangle.
\end{eqnarray}
Hence, the chiral anomaly which we have considered in the previous section
still exists when we include the Coulomb interaction into the analysis~\cite{Iso}.

\section{effects of external charges}\label{sec:eff-external}

We have so far considered the Coulomb interaction between the 
internal charges. However, more important problem may be the following:
when we put external charges on a nanotorus,
how the vacuum structure change ?
A fixed external charge may corresponds to an
electrical contact or an impurity on a torus.
In this section we put external charges on a torus and study the 
vacuum structure. Especially we analyze an effect of external charges on the
chiral charge, 
the potential behavior between a pair of external charges and 
the `charge screening' effect. The charge screening means that 
the internal charge density is induced by the external charges so that
the external charges are screened.

\begin{figure}
\begin{center}
\includegraphics[scale=0.4]{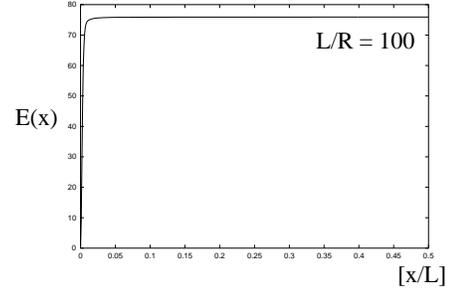}
\end{center}
\caption{Distance dependence of the energy caused by two fixed 
external charges. The energy value is labeled in the unit of 
$\frac{2\pi \hbar v_F}{L}$. The distance between two external charges 
is scaled by $L$. Here we take $\frac{L}{R} = 10^2$ in 
eq.(\ref{eq:Fourier-coulomb}).}
\label{fig:externalcharge}
\end{figure}

We set two unit external charges on a torus.
One has a unit charge $e$ placed at $x_0$ and the other 
has an opposite charge $-e$ at $y_0$.
\begin{eqnarray}
J^0_{ex}(x) = &&
\delta(x-x_0) - \delta(x-y_0) \nonumber \\
= && \frac{1}{L} \sum_{n \in Z} j^n_{ex} e^{-i\frac{2\pi nx}{L}},
\end{eqnarray}
where 
\begin{eqnarray}
j^n_{ex} = e^{i \frac{2\pi n x_0}{L}} - e^{i \frac{2\pi n y_0}{L}}.
\end{eqnarray}
We consider the Coulomb interaction with the following charge density
which is a sum of the internal charges and the external charges:
\begin{eqnarray}
&&
J^0(x)+ J^0_{ex}(x) = \nonumber
\\
&&
\ \ \ \ \ \sum_{n \in Z} \left(
(j_L^n)^\dagger + j_R^n + (j^n_{ex})^*
\right) \frac{1}{L} e^{+i \frac{2\pi nx}{L}}.
\end{eqnarray}
After some calculations we get 
\begin{eqnarray}
H_n(J_{ex}) \equiv &&
E_n \Big[ 
\left( (\tilde{j}_L^n)^\dagger + \gamma_n (j^n_{ex})^* \right)
\left( \tilde{j}_L^n + \gamma_n j^n_{ex} \right) \nonumber \\
&&
+ \left( (\tilde{j}_R^n)^\dagger + \gamma_n j^n_{ex} \right)
\left( \tilde{j}_R^n + \gamma_n (j^n_{ex})^* \right) 
+ n \Big] 
\nonumber \\
&&- \frac{2\pi \hbar v_F}{L}n + \frac{\beta_n}{E_n^2} 
\left( \frac{2\pi \hbar v_F}{L} \right)^2 (j_{ex}^n)^* j_{ex}^n ,
\end{eqnarray}
where $\gamma_n = \sinh 2t_n (\cosh t_n - \sinh t_n ) $ and 
$\beta_n = \frac{e^2}{4\pi L}V(n)$.
It is easy to find conditions of the vacuum 
in the presence of the external charges,
\begin{eqnarray}
&&
\left( \tilde{j}_L^n + \gamma_n j^n_{ex} \right) 
|\widetilde{vac} ;J^0_{ex} \rangle
= 0 , 
\\
&&
\left( \tilde{j}_R^n + \gamma_n (j^n_{ex})^* \right)
|\widetilde{vac} ;J^0_{ex} \rangle = 0,  \ \ n > 0.
\end{eqnarray}
First we can easily show that the chiral charge on this new vacuum
does not change even in the presence of external charges,
\begin{eqnarray}
\langle \widetilde{vac} ;J^0_{ex}| J^5(x) |\widetilde{vac} ;J^0_{ex} \rangle
= \frac{\langle Q_5 \rangle}{L}.
\end{eqnarray}
So, the current by the chiral anomaly is not affected by the charged impurities.
Second, we estimate the energy change due to the presence of the external charges:
\begin{eqnarray}
&&
E(x_0-y_0) = \langle \widetilde{vac}; J^0_{ex}|H(J_{ex})|\widetilde{vac};J^0_{ex} 
\rangle - \langle \widetilde{vac} |H| \widetilde{vac} \rangle \nonumber \\
&&=
\sum_{n >0} \frac{\beta_n}{E_n^2} 
\left( \frac{2\pi \hbar v_F}{L} \right)^2 (j_{ex}^n)^* j_{ex}^n
\nonumber
\\
&&
= \frac{2\pi \hbar v_F}{L} \sum_{n >0}
\frac{\frac{\alpha}{\pi}\frac{c}{v_F}V(n)}{1 + \frac{\alpha}{\pi}\frac{c}{v_F} V(n)}
\left( 1- \cos \frac{2\pi n (x_0-y_0)}{L} \right),
\label{eq:bindene}
\end{eqnarray}
where $H(J_{ex})$ is the Hamiltonian with the external charges($J_{ex}$).
Fig.\ref{fig:externalcharge} shows
the potential energy (eq.(\ref{eq:bindene})) as a function of 
the distance between the two external charges.

\begin{figure}
\begin{center}
\includegraphics[scale=0.4]{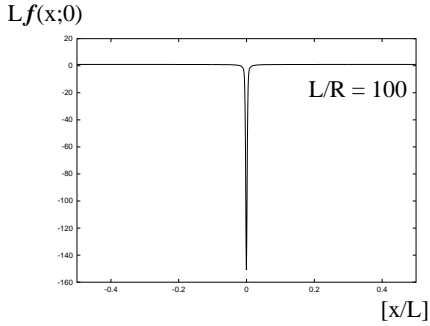}
\end{center}
\caption{The internal charge density modulation around a 
unit charge $e$ placed at origin: $f(x;0)$. The charge density value 
and the distance from the charge are labeled in the unit of $L$ and $\frac{1}{L}$ 
respectively.
Here we take $\frac{L}{R} = 10^2$ in 
eq.(\ref{eq:Fourier-coulomb}).}
\label{fig:internalcharge}
\end{figure}

We see that the effect of the `charge screening' on
the external charge can not be ignored in a metallic nanotorus.
Because the potential energy is now shown to be short-ranged. 
This means that some internal charges are 
influenced by the external charges, and external charges are screened.
To confirm this, 
we also compute the induced internal 
charges distribution,
\begin{eqnarray}
\langle \widetilde{vac} ;J^0_{ex} |J^0(x)| \widetilde{vac} ;J^0_{ex} \rangle 
= f(x;x_0) - f(x;y_0),
\end{eqnarray}
where
\begin{eqnarray}
f(x;x_0) = 
- \frac{2}{L} 
\sum_{n >0} \frac{\frac{\alpha}{\pi}\frac{c}{v_F}V(n)}{1 + 
\frac{\alpha}{\pi}\frac{c}{v_F}V(n)}
\cos \frac{2\pi n (x-x_0)}{L}. 
\label{induced-ch}
\end{eqnarray}
This function is displayed in Fig.\ref{fig:internalcharge}.

We should stress here that the above analysis on the charging energy (\ref{eq:bindene})
and screening (\ref{induced-ch}) are not complete because the charge density 
($J^0(x)$) in the Coulomb interaction consist of one massless fermion in our analysis.
A CNT have four independent fermions due to the two Fermi points(\ref{2fermi-points}) and 
spin degrees of freedom. However, straight forward extension of our analysis shows that
the conclusions in this paper hardly changes~\cite{KS}.

\section{conclusion and discussion}\label{sec:discuss}
In this paper we analyzed the carbon nanotorus 
and discussed the quantum mechanical vacuum structure
of a metallic torus.
The significance of our present work may be put as follows. 

We used the quantum field theory to analyze the vacuum structure of 
a metallic nanotorus.
We pointed out that the chiral anomaly in 1+1 dimensions should
be observed in the form of specific magnetic field dependence to the current. 
This current is the same as the persistent current. 
It is certain that the persistent current can be understood
in the light of the chiral anomaly in 1+1 dimensions.
We also clarified the 
vacuum state including the Coulomb interaction and discussed the effect
of `charge screening' on the external charges.
It was found that the chiral anomaly is not affected by the charged impurities and 
the charge screening effect actually occurs.

Although we have analyzed a nanotorus in the present paper,
we can regard the torus as a tube locally provided the circumferential
length of the torus is large.
Hence, we may apply the results obtained in the analysis of the torus 
to the study of a carbon nanotube.
To take an example, it is reasonable to suppose that the screening effect 
is also present in a tube like in a nanotorus.

What we do not consider in this paper are 
effects of finite temperature( $> \frac{2\pi\hbar v_F}{L}$ )
on chiral anomaly and screening of external charges.
In finite temperature, in addition to the 
electron field, phonon field would come into play~\cite{SS}.
This phonon field may be vital to understand the electrical behavior 
of the carbon nanotube at finite temperature.
We would like to make a quantitative analysis of finite
temperature effects in a future report.

\section{Acknowledgments}
The author would like to thank M. Hashimoto
for various discussions on the subject.
This work is supported by a fellowship of the Japan 
Society of the Promotion of Science.




\begin{thebibliography}{99}
\bibitem{Iijima}
S. Iijima, Nature, {\bf 354}, 56 (1991).
\bibitem{RS}
R. Saito, G. Dresselhaus and M.S. Dresselhaus, \\
{\it Physical Properties of Carbon Nanotubes} \/(1998) \\
Imperial College Press, London. 
\bibitem{KT}
{\it The Science and Technology of Carbon Nanotubes} \/(1999) \\
Ed. by K. Tanaka, T. Yamabe and K. Fukui, Elsevier Oxford. 
\bibitem{Dek}
C. Dekker, Phys. Today {\bf 52}, No. 5, 22 (1999).
\bibitem{JWM}
J.W. Mintmire, B.I. Dunlap and C.T. White, 
Phys. Rev. Lett. {\bf 68}, 631 (1992);
N. Hamada, S. Sawada and A. Oshiyama, 
Phys. Rev. Lett. {\bf 68},1579 (1992);
R. Saito et al.,
Appl. Phys. Lett. {\bf 60}, 2204 (1992).
\bibitem{WV}
J. W. G. Wild\"oer et al., Nature, {\bf 391}, 59 (1998);
T. W. Odom et al., Nature, {\bf 391}, 62 (1998).
\bibitem{cc}
J. Liu et al. Nature, {\bf 385}, 781 (1997).
\bibitem{Bell}
J.S. Bell and R. Jackiw, Nuovo Cim. {\bf 60}, 47 (1969);
S. Adler, Phys. Rev. {\bf 177}, 2426(1969).
\bibitem{Sch}
J. Schwinger, Phys. Rev. {\bf 128}, 2425 (1962);
J. Lowenstein and A. Swieca, Ann. Phys. {\bf 68}, 172 (1971).
\bibitem{Man}
The massless Schwinger model (quantum electrodynamics with massless fermion in one 
spatial dimension) is first solved in the Hamiltonian formalism by,
N. S. Manton, Ann. Phys. {\bf 159}, 220 (1985).
\bibitem{Iso}
Most of the detail about the massless Schwinger model can be found in,
S. Iso and H. Murayama, Prog. Theor. Phys. {\bf 84}, 142 (1990).
\bibitem{Jack}
R. Jackiw, in: Lectures on current algebra and its applications, eds. 
S. Treiman, R. Jackiw and D. Gross, Prinston Univ. Press, Prinston, (1972).
\bibitem{NN}
H.B. Nielsen and M. Ninomiya, Phys. Lett. {\bf 130B}, 389 (1983);
I. Krive and A. Rozhavsky, Phys. Lett. {\bf 113A}, 313 (1983);
M. Stone and F. Gaitan, Ann. Phys. (NY) {\bf 178}, 89 (1987).
\bibitem{persist}
The persistent current in toroidal carbon nanotubes is analyzed by,
R.C. Haddon, Nature, {\bf 388}, 31 (1997);
M.F. Lin, Phys. Rev. B. {\bf 57}, 6731 (1998);
A.A. Odintsov, W. Smit and H. Yoshioka, Europhys. Lett., {\bf 45}, 598 (1999).
\bibitem{JG}
J. Gonz\'alez, F. Guinea and M.A.H. Vozmediano,
Nucl. Phys. {\bf B406}, 771 (1993).
\bibitem{Multi}
Y. -K. Kwon and D. Tom\'anek,
Phys. Rev. B. {\bf 58}, R16001 (1998).
\bibitem{Kim}
P. Kim et al., Phys. Rev. Lett. {\bf 82}, 1225 (1999).
An experimental result suggests that curvature-induced 
hybridization is only a small perturbation for the (13,7) tube.
\bibitem{CBJ}
W. Clauss, D. J. Bergeron and A. T. Johnson, 
Phys. Rev. B. {\bf 58}, R4266 (1998).
\bibitem{note}
It should be noted that to obtain eq.(\ref{eq:pert}) we have to 
perturb the Hamiltonian around the points $a\tilde{k}_1 - aK_1$
and $a\tilde{k}_2 + aK_1$. The wave vectors $K_i$ satisfy the conditions,
$ K_i \cdot T_j = 2\pi \delta_{ij}$.
The vectors 
$a\tilde{k}\pm aK_1$ and $a\tilde{k}$ represent the same state since any 
two wave vectors congruent by $aK_i$ are just different labels of the
same state.
\bibitem{weyl}
D. P. DiVincenzo and E. J. Mele, Phys. Rev. B {\bf 29}, 1685 (1984);
H. Ajiki and T. Ando, Solid. State. Commun. {\bf 102}, 135 (1997);
T. Ando and T. Nakanishi, J. Phys. Soc. Jpn. {\bf 67}, 1704 (1998).
\bibitem{betaE}
In order to get the unnormalized energy we should 
multiply the Fermi velocity $v_F(\equiv |\beta|) $ by $\epsilon_n $. We estimate the order of 
magnitude of energy spectrum using $v_F \sim \frac{c}{400}$.
Here we use $\gamma = 2.5 $[eV] and $a = 1.42 $[\AA],
$\beta \epsilon_n \sim 5 (eV) \left[\frac{\AA}{L}\right] 2\pi 
\left(n - N \right)$.
\bibitem{EG}
R. Egger and A.O. Gogolin, Phys. Rev. Lett., {\bf 79}, 5082 (1997).
\bibitem{KS}
K. Sasaki, work in progress.
\bibitem{SS}
B. Sakita and K. Shizuya, 
Phys. Rev. B. {\bf 42}, 5586 (1990).


\end{thebibliography}
\end{document}